\documentclass[referee]{raa}            

\usepackage{graphicx,times}             
\usepackage{natbib}
\usepackage{ulem}
\usepackage{color}
\usepackage{amssymb,amsmath}
\bibpunct{(}{)}{;}{a}{}{,}

\usepackage[a4paper=true,dvipdfm=true,pagebackref=true]{hyperref}
\hypersetup{colorlinks = true, linkcolor = green, anchorcolor = red, citecolor = blue, filecolor = red, pagecolor = red, urlcolor = red}

\begin{document}
   \title{A PRESTO-based Parallel Pulsar Search Pipeline Used for FAST Drift Scan Data}


   \volnopage{Vol.0 (20xx) No.0, 000--000}      
   \setcounter{page}{1}          

   \author{Qiuyu Yu
      \inst{1}
   \and Zhichen Pan
      \inst{2}
   \and Lei Qian
      \inst{2}
   \and Shen Wang
      \inst{2}
   \and Youling Yue
      \inst{2}
   \and Menglin Huang
      \inst{2}
   \and Qiaoli Hao
      \inst{2}
   \and Shanping You
      \inst{2}
   \and Bo Peng
      \inst{3}
   \and Yan Zhu
      \inst{4}
   \and Lei Zhang
      \inst{2}
   \and Zhijie Liu
      \inst{1}
   }

   \institute{Guizhou Normal University, Key Laboratory of Information and Computing Science, Guiyang, 550001, China;{\it liuzj@gznu.edu.cn}\\
        \and
             National Astronomical Observatories, Chinese Academy of Sciences,
             Beijing 100012, China; {\it panzc@bao.ac.cn}\\
        \and
             School of Information Engineering, Southwest University of Science and Technology Mianyang, 621010, China;\\
        \and
             School of physics, university of electronic science and technology of china, sichua, 610054, China;\\
\vs\no
   {\small Received~~20xx month day; accepted~~20xx~~month day}}

\abstract{ We developed a pulsar search pipeline based on PRESTO (PulsaR Exploration and Search Toolkit). This pipeline simply runs dedispersion, FFT (Fast Fourier Transformation), and acceleration search in process-level parallel to shorten the processing time. With two parallel strategies, the pipeline can highly shorten the processing time in both the normal searches or acceleration searches. This pipeline was first tested with PMPS (Parkes Multibeam Pulsar Survery) data and discovered two new faint pulsars. Then, it was successfully used in processing the FAST (Five-hundred-meter Aperture Spherical radio Telescope) drift scan data with tens of new pulsar discoveries up to now. The pipeline is only CPU-based and can be easily and quickly deployed in computing nodes for testing purposes or data processes.
}

   \authorrunning{Q. Yu, Z. Pan. L. Qian et al}
   \titlerunning{A PRESTO-based Parallel Pulsar Search Pipeline Used for FAST Drift Scan Data}  

   \maketitle

%
%
\section{Introduction}
\label{sect:intro}

  A typical pulsar search process consists
          1) selecting and marking the Radio Frequency Interferences (RFIs),
          2) dedispersion,
          3) transforming the time domain data to a frequency domain signal, for example, using Fast Fourier Transform (FFT),
          4) searching for periodic signals,
          5) and finding pulsar candidates from search results.
  It is always one of the goals of pulsar search software to accelerate these steps in order to shorten the data processing time.
  One of the basic ideas is finding a suitable pulsar searching plan.
  For example,
  reducing the number of dispersion trails is equivalent to reducing the total computational effort and thus reducing the data processing time.
   Hardware like GPUs (Graphic Processing Units), also accelerates pulsar searches (e.g., the code peasoup\footnote{https://github.com/ewanbarr/peasoup}, Barr et al. in prep).
  In dedispersion, a GPU/multicore-based code can easily reduce the data processing time to 50\% of the original value or even shorter (\citealt{Sclocco+etal+2016}).
  Using multi-core CPUs (Central Process Units) and GPUs, the coherent dedispersion pipeline of Giant Metrewave Radio Telescope (GMRT) is better than real-time (\citealt{De+Gupta+2016}).
  For the LOw Frequency ARray(LOFAR), GPUs speed-up de-dispersion over 2.5 times faster than previous Fermi code with rough estimation (\citealt{Serylak+etal+2013}).
  For periodic signal searches, GPU-based acceleration search codes (e.g. Luo et al. in prep) also reduce the searching time.
  For the Sourthern High Time Resolution Universe (HTRU) pulsar survey,
  the GPU-accelerated dedispersion and periodicity search codes are nearly 50 times faster than the previously used pipeline (\citealt{Morello+etal+2019}).
  The Fourier Domain Acceleration Search (FDAS) of the Square Kilometre Array (SKA) has achieved better than real-time performance (\citealt{Dimoudi+Armour+2017}).
  For single pulse search, GPU-based heimdall\footnote{https://sourceforge.net/projects/heimdall-astro/} (\citealt{Barsdell+etal+2010}),
  and ROACH2-based (ROACH2\footnote{https://casper.ssl.berkeley.edu/wiki/ROACH2}) hardware, or even a CPU-based code (Lee et al. in prep) can process data in realtime.
  As an example at the Green Bank Telescope (GBT),
  GPU-accelerated codes are now available to optimize this dedispersion task, and to search for transient pulsed radio emission (\citealt{Walsh+Lynch+2018}).

  As a special situation,
  for drift scan survey (e.g., CRAFTS\footnote{http://crafts.bao.ac.cn}) with FAST (Five-hundred-meter Aperture Spherical radio Telescope, \citealt{Nan+etal+2011, Jiang+etal+2019}),
  we are going to process many small files (e.g., less than 100 MB) and wish the search can be done in realtime.
  This is a challenge for both computing and disk input/output.
  For example, at the frequency of $\rm \sim$300 MHz, the beam size of FAST ($\rm \sim$ 13 arcminutes) corresponds to $\rm \sim$52 seconds transit time.
  With 50\% beam overlap, we have to process one file within 26 seconds on average in order to process the data in realtime.
  If the number of channels is 256, the sampling time is 0.2 millisecond, and it is 8-bit sampling, the size of a data file is $\rm \sim$60 MB. as typical pulsar search requires reading and writing frequently.

  GPUs are suitable for computing rather than reading or writing.
  Thus, a GPU-based code for processing small files may hardly reach high performance.
  Secondly, we do not have GPUs for processing FAST drift scan data during the first two years of FAST commissioning time.
  In order to search for new pulsars, we need a CPU-based pulsar search pipeline.
  In addition, with many pulsar search examples in the package,
  PRESTO  (\citealt{Ransom+2001, Ransom+etal+2002, Ransom+etal+2003}) is an open source code that has already discovered many pulsars, is updated and widely used till now (\citealt{Swiggum+Gentile+2018}).
  So, we started our work with PRESTO.

  In this paper, we present our PRESTO-based pipeline for FAST drift scan pulsar search data.
  The details of the pipeline are presented in Section 2.
  The testing results on both the PMPS (Parkes Multibeam Pulsar Survey, \citealt{Manchester+etal+2001}) data and the real FAST data, are given in Section 3.
  In Section 4 and 5, there are data processing results and discussion, respectively.
  Section 6 is the conclusion.

\section{The Pipeline}
\label{sect:The Pipeline}
  The pulsar search relies on CPU, memory, and disk Input/Output operations Per Second (IOPS) to some degree.
  For large data files (e.g. with size $\ge$1 GB), the processing time of dedispersion, FFT, or the search phase depends largely on the processing power of the CPU.
  If acceleration search or phase modulation search is used, a large size of memory may be required.
  For small files, normal pulsar search pipelines based on PRESTO or SigProc\footnote{http://sigproc.sourceforge.net and https://github.com/SixByNine/sigproc} (\citealt{Lorimer+2011}) require writing and reading files frequently.
  As an improvement, some GPU-based codes, such as $peasoup$,
  provide a one-step pulsar search with only one command (can be very long) and with the candidate list the only output file.
  For FAST drift scan pulsar survey, we use a cluster with 20 nodes (480 cores in total, Intel E5 2680 v3 2.5 GHz CPU). For data processing, we prepared a CPU-based pulsar search pipeline with PRESTO routines.

  A typical PRESTO-based pulsar search pipeline (for example, PRESTO provides several examples: GBNCC\_search.py, full\_analysis.py, and PALFA\_presto\_search.py) processes the data with the following steps:

  1, finding and masking RFIs (using {\bf rfifind});

  2, dedispersing the data into time series with many DM values (using {\bf prepdata} or {\bf prepsubband});

  3, transforming the time series data to the frequency domain data (using {\bf realfft}). This step is not necessary, while {\bf accelsearch} can do this transformation, too;

  4, searching for periodic signals from the frequency domain data (using {\bf accelsearch} );

  5, sifting the search result and obtaining candidates (using {\bf ACCEL\_sift.py});

  6, folding the data with the periods and DM of the candidates (using {\bf prepfold});

  7, searching for single pulses from the time series {\bf (using single\_pulse\_search.py *.dat)};

  For steps 2, 3, 4, and 6,
  the corresponding routines ({\bf prepdata} or {\bf prepsubband}, {\bf realfft}, {\bf accelsearch}, and {\bf prepfold}) or same ({\bf realfft}) will be run for multiple times with different options.
  Those routines can run in parallel to save the time.
  Along with this idea,
  we use Linux shell scripts, C language, and python code to re-organise the PRESTO-based pulsar search pipeline,
  so that the pipeline starts multiple processes to dedisperse the data, FFT, search, and fold at the same time, thus reducing the total time cost.
  For dedispersing, we use {\bf prepdata}, because this is the routine used for dedispersion.
  In fact, it is not always a good choice.
  We will discuss this in Section 5.
  The name of the pipeline is RPPPS\footnote{https://github.com/qianlivan/RPPPS},
  which is the abbreviation for Re-analysing Pipeline for Parkes Pulsar Survey,
  since the pipeline was originally designed for reprocessing PMPS data.

  There are two key values for RPPPS.
  One key parameter for RPPPS is the maximum number of processes that run simultaneously (called parallel number hereafter).
  The other one is used to control how the code runs in parallel (called parallel version hereafter).
  The pipeline may run routines round by round (hereafter V2),
  or one by one according to the number of running processes (hereafter V3)\footnote{V1 is for the non-parallel testing pipeline and will never be used again}
  For example,
  if the parallel number is 128, with V2 option,
  the code will start 128 processes (for dedispersing, FFT, and acceleration search in serial) simultaneously,
  count until all the processes are finished, and start another 128 processes.
  With V3 option, the code will start one process and check how many processes are running in total.
  If the number of processes is less than 128, it will start one more.
  If the number of processes is equal or larger than 128, it will wait for several seconds and check the number of processes again.

  With such parallel strategies, the PRESTO-based pulsar search pipeline can be accelerated a lot.

\section{Test and results}
\label{sect:Test and results}
\subsection{Test hardware platform and data files}

  To test our pipeline,
  we performed tests on two computers with different configurations.
  The configurations of these computers are

  1) Two Intel Gold 6130 $\times$ 2, 2.1 GHz, 32 cores in total, 192 GB DDR4 RAM

  2) Two Intel E5-2680 v3 $\times$ 2, 2.5 GHz, 24 cores in total, 512 GB DDR3 RAM, which is the configuration of the computing nodes for FAST drift scan pulsar search.

  As for the data files used for the test,
  we randomly selected a PMPS file from the public Parkes database, PM0064\underline{ }00311.sf, as the test data file.
  In order to avoid the test results being affected by different pulsar searching results from different files, all tests use the same file.
  Another test file is a real observational file from FAST (also randomly selected).
  The details of these two files are shown in Table \ref{test_files}.

\begin{table}
\begin{center}
\caption[]{test files details}\label{test_files}
 \begin{tabular}{clcl}
  \hline\noalign{\smallskip}
                & PMPS File & FAST File \\
  \hline\noalign{\smallskip}
  File Name     &  PM0064\underline{ }00311.sf & FAST\underline{ }test.fits              \\
  File Size     &  $\rm \sim$100 MB  &  $\rm \sim$51 MB \\
  Number of Channels  &  96  &  200  \\
  Bandwidth     &   288 MHz  &  50 MHz \\
  Band          &   1231.5-1516.5 MHz         & 290-340 MHz \\
  Width of Channels   &  3 MHz   & 0.25 MHz  \\
  Observing time  & 2100.224 s  &  52.4288 s \\

  \noalign{\smallskip}\hline

\end{tabular}
\end{center}
\end{table}

  Another setting related to the amount of calculation is the DM values for de-dispersion.
  For the PMPS file, we choose approximate 0 to 3500 pc cm$^{-3}$ as the dispersion range.
  For FAST data, the DM range is approximate 0 to 800 pc cm$^{-3}$.
  Then the {\bf DDplan.py} program in PRESTO was used to generate a dedispersion scheme.
  The dedispersion plan for PMPS data and FAST data are shown in the Table \ref{PMPS_dm} and \ref{FAST_dm}.

\begin{table*}
	\centering
	\caption{Dedispersion Scheme Details for PMPS Data}
	\label{tab:example_table}
	\begin{tabular}{lccccc} 
	 \hline
    No.  &  DM Start      &   DM End        &  DM Step         &   Down Sample Factor  & Number of DMs  \\
		 & (pc cm$^{-3}$) & (pc cm$^{-3}$)) & (pc cm$^{-3}$)   &                       &                 \\
     \hline
		1 & 0      & 114.0  & 0.5  & 1   & 228 \\
		2 & 114.0  & 266.0  & 1.0  & 2   & 152 \\
        3 & 266.0  & 476.0  & 2.0  & 4   & 105  \\
        4 & 476.0  & 856.0  & 5.0  & 8   & 76  \\
        5 & 856.0  & 2376.0 & 10.0 & 16  & 152 \\
        6 & 2376.0 & 3896.0 & 20.0 & 32  & 76  \\
		\hline
\label{PMPS_dm}
	\end{tabular}
\end{table*}

\begin{table*}
	\centering
	\caption{Dedispersion Scheme Details for FAST Ultra-wideband Drift Scan Data}
	\label{tab:example_table}
	\begin{tabular}{lccccc} 
	 \hline
    No.  &  DM Start      &   DM End        &  DM Step         &   Down Sample Factor  & Number of DMs  \\
		 & (pc cm$^{-3}$) & (pc cm$^{-3}$)) & (pc cm$^{-3}$)   &                       &                 \\
     \hline
		1 & 0      & 31.5   & 0.1  & 8    & 315 \\
		2 & 31.5   & 62.9   & 0.2  & 16   & 157 \\
        3 & 62.9   & 139.4  & 0.5  & 32   & 153 \\
        4 & 139.4  & 278.4  & 1.0  & 64   & 139 \\
        5 & 278.4  & 556.4  & 2.0  & 128  & 139 \\
        6 & 556.4  & 784.4  & 3.0  & 128  & 76  \\
		\hline
\label{FAST_dm}
	\end{tabular}
\end{table*}

\subsection{Differences in Storages, Hyper-Threading, and Turbo Boost Technique}

  Currently used storage environments include mechanical HDDs (hard disk drivers) and SSDs (solid-state hard disks), as well as network file systems.
  In addition, the computer's memory (RAM) can be divided into storage space for temporary usage,
  e.g., the /dev/shm partition under Linux is a virtual storage from memory,
  which has a much faster read/write speed than HDDS or SSDs.
  In order to test if different storage environments will affect the speed of data processing,
  we selected the HDDs and the /dev/shm partition of Linux, for testing .
  We processed the same PMPS data file in the memory virtual partition and the HDD, respectively.
  In the test, we used the default settings of the server, which are enabling TBT (Turbo Boost Technology) and disabling HT (hyper-threading).
  In order to cause more IOPs pressures,
  we use V2 as the parallel version, and set zmax value for {\bf accelsearch} and parallel number to be 0 and 128/1024, respectively.
  We repeated 10 times for each storage.
  The platform we used is the first one which is mentioned in Section 2.1.

  The result is that whatever the data processing was running on HDD or the /dev/shm folder,
  the time cost is between 250 seconds and 270 seconds when the number of parallel process was set to be 128 or 1024,
  though in fact there is only 32 real CPU cores in total.
  So, a HDD can support running such a parallel pulsar search pipeline.
  the reason for the HDD and SDD having similar performance is  probably because of Linux filesystem caching on these machines with lots of RAM.

  The HT feature can double the core number in the operating system.
  In common situation, e.g., a desktop for daily usage, the HT is turned on,
  while it is suggested to be turned off for high performance computing.
  Since we are using a simple parallel approach, it is necessary to see if enabling HT will affect the data processing time or not.
  TBT is a feature that the CPU automatically raises its clock frequency.
  According to this principle, it seems that enabling this function should be more conducive to shortening the data processing time.

  Table \ref{HT_TBT} shows the average time cost whether the TBT and/or HT will be used or not.
  We used the V2 parallel version and set the parallel number to be 128 for this test.
  In order to prevent the errors, we ran 10 times and obtain one averaged value.
  The numbers in bold is the shortest time cost.
  It is obvious that for non-acceleration search, TBT and HT should be enabled,
  while for acceleration search, TBT should be enabled and HT should be disabled.

  According to the tests above,
  we will enable TBT and HT, and use the shared network file system for further tests on the computing node which would be used for FAST drift scan pulsar search.

\begin{table*}
	\centering
	\caption{HT and TBT Affect Time Cost}
	\begin{tabular}{lccccc} 
	 \hline
    TBT.        &  HT       &   ZMAX  & Average Time Cost (s)  \\
     \hline
		Enable  & Enable    &    0     & \textbf{226.5}  \\
		        &           &    50    & 431.1  \\
                & Disable   &    0     & 267.2  \\
                &           &    50    & \textbf{424.8}  \\
    \hline
        Disable & Enable    &    0     & 297.3  \\
                &           &    50    & 465.9  \\
                & Disable   &    0     & 304.0  \\
                &           &    50    & 491.7  \\
		\hline
\label{HT_TBT}
	\end{tabular}
\end{table*}

\subsection{The relationship between the number of processes and processing time}

  The time cost of our pipeline may be affected by the following rules.

  1) when the number of processes running in parallel is less than the number of CPU cores, the CPU usage is not 100\% used.
     The time cost should be longer than expected.

  2) when the number of processes running in parallel is equal or several times larger than the number of CPU cores,
     the total computation time is close to the shortest one.

  3) when the number of processes running in parallel is much larger than the number of CPU cores,
  the total time cost changes little due to that the computer can't provide more computing resources.
  What's more, this will cause the computer to stop responding and thus the time cost will increase.

  In order to detect these situation,
  we processed the same FAST file with different parallel process numbers on the computing nodes (the second one mentioned in Section 2.1),
  to see if the processing time was shortened when the number of parallel processes is increasing.
  The number of processes starts at 8, in steps of 8, until 1024.

\begin{figure}
  \centering
    \includegraphics[width=80mm]{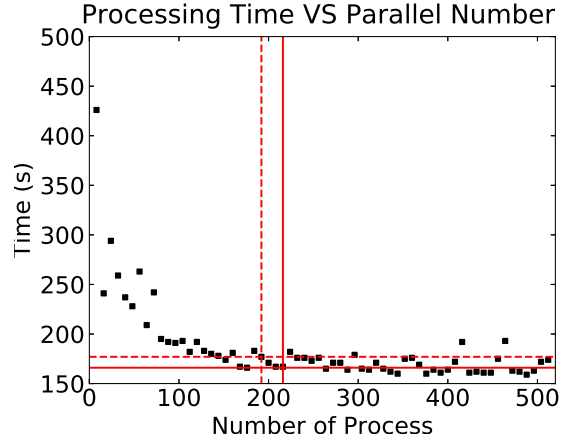}
      \caption{Processing time reduces when the parallel number increases.
               Vertical and horizontal dashed red lines show the parallel number 192 and processing time 177 second, respectively.
               Vertical and horizontal red lines show the parallel number 216 and processing time 166 second, respectively. }
  \label{process}
\end{figure}

  Figure \ref{process} shows the time cost decreases when the number of process increases .
  We finally selected 192 (177 seconds) and/or 216 (166 seconds) as the number of process for the FAST drift scan pulsar search.
  In real situation, the total time cost for processing FAST drift scan data with the parallel number 192 or 216 are almost same.

\section{Results}
\label{sect:Results}
  We then applied the RPPPS as the pulsar search pipeline to find possible new pulsars from archived PMPS data and the FAST drift scan data.
  This resulted in in two new pulsar discoveries from PMPS data and tens of new pulsars discovered from FAST data.

\subsection{PMPS Data Reprocessing}

  We processed $\rm \sim$16\% of the PMPS dataset (8240 files) for pulsar search testing, aiming find new faint pulsars missed by previous processing.
  It costs $\rm \sim$14 days to process all the files with the acceleration search in which the zmax value was 50, and resulted in 410535 candidates.
  All those candidates were viewed by human beings and 2238 good candidates were finally selected.
  Among them, the sixteen best candidates were then re-observed by Parkes in the April of 2018 and July of 2019.
  Two were confirmed to be new pulsars (middle panels of Figure \ref{cands}).
  They are very faint pulsars with medium DM values.
  The pulsar J1900-04 is relatively bright in both PMPS data and our confirmation observation data.
  The pulsar J1808-12 is so faint that we can not search and find it in our confirmation data.
  If we do not know its period and DM from PMPS data reprocessing, we can't find it.
  The details of discovering these two new pulsars will be in another paper (Pan et al. in prep).

  The PMPS data reprocessing indicates that the pipeline has a good performance on pulsar search and is potentially good at finding faint pulsars.

\subsection{FAST Drift Scan}

  From August 2017, this pipeline was used to process FAST drift scan pulsar search data.
  The ultra wideband receiver covers a band of 290-1760 MHz.
  We used the 290-340 MHz band (200 channels) for pulsar search with RPPPS.
  The characteristics of every file are the same as the one we used to test the pipeline (see details in Table \ref{test_files}).
  With the cluster mentioned before, processing 12-hour drift scan data will cost approximately the same time, which means that the data can be processed in realtime.

\begin{figure}
  \centering
    \includegraphics[width=80mm]{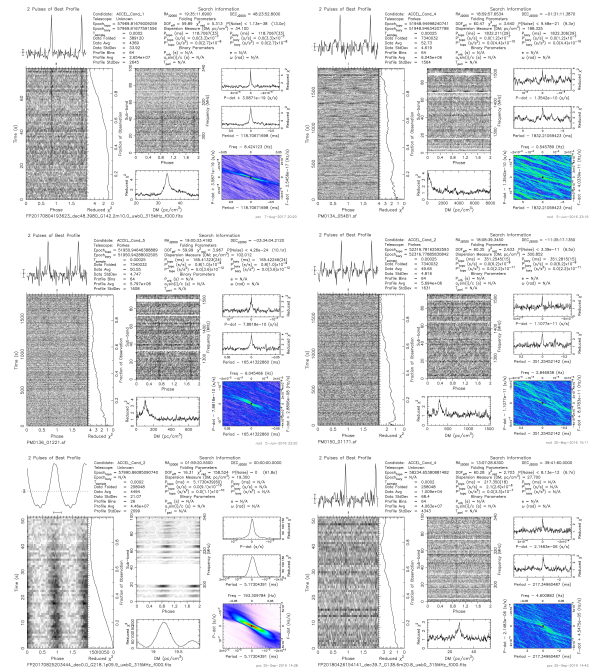}
      \caption{$Upper Left$: the discovery plot of the first FAST pulsar candidates (now confirmed), designated as C1;
               $Upper Right$: the PMPS re-detection of the first FAST confirmed pulsar (J1900-0134), see Qian et al. (2019) for the plot from corresponding FAST data;
               $Middle Left and Right$: the two confirmed new pulsars from reprocessing of PMPS data, discovery details will be in Pan et al. (in prep);
               The left one is J1900-04, the right one is J1808-12;
               $Lower left$: the first FAST MSP candidates (now confirmed), discovered by RPPPS, designated as C8;
               $Lower Right$: a very faint FAST candidate discovered by RPPPS, designated as C59.}
  \label{cands}
\end{figure}

  With RPPPS, the first FAST pulsar candidate was discovered on August 6$^{th}$, 2017, on the FAST's August 4$^{th}$ drift scan data.
  Figure \ref{cands} (upper left) show the discovery plot.
  It is a relative strong pulsar and no doubt that it should be found.
  This candidates was then confirmed by FAST in August 6th.
  This is also the first real pulsar discovery by RPPPS.

  As a cross check between Parkes data and FAST data,
  the first confirmed FAST pulsar (J1900-0134, \citealt{Qian+etal+2019}) can also be re-discovered from PMPS data by RPPPS.
  Figure \ref{cands} (upper right) shows the re-discovery of this pulsar by RPPPS from the PMPS data.
  The plot from corresponding FAST data can be found in \citealt{Qian+etal+2019}.
  This proves that the RPPPS can be suitable to process data from different telescopes, and find new bright or faint pulsars.

  The first MSP (millisecond pulsar) candidate was also detected by RPPPS (lower left panel of Figure \ref{cands}).
  Approximately one third of the band was removed to reduce the RFIs, yet this MSP is still very bright and shows scintillation features.
  The RPPPS also found faint candidates from the drift scan data, such as the one in the lower right panel of Figure \ref{cands}.

  Till now, more than 40 candidates have been found with RPPPS, including high DM pulsars, millisecond pulsars, and binary pulsars.
  Most of them are already confirmed to be new pulsars.

\section{Discussions}
\label{Discussions}

\subsection{Realtime Data Processing:}
  According to the test result, one computing node can process approximately 26 seconds of observation data in $\rm \sim$170 seconds.
  Within 20 nodes, processing one-night (10-12 hours) of drift scan data should be finished within $\rm \sim$4 hours (33\% of the observation time).
  In fact, the data processing time reaches $\rm \sim$80\% of the observing time.
  This is mainly due to the bottleneck from the shared network file system IOPS.
  The computing nodes have very limited disk space thus we can't run the whole search on local disks.
  Otherwise,
  we may firstly copy the data file to the local disk of the computing node and move the search result back to the shared file system after data process finished,
  and thus accelerate the pulsar search more.

  In our previous test, the time costs when using different local storages are almost same.
  When using clusters and when many computing nodes were reading and writing at the same time, the shared network file system should be high performance one.

\subsection{Using HT:}
  After running for several months, we upgraded the cluster and enabled HT for all the computing nodes,
  because after optimizing by Sugon, enabling HT will shorten the time cost more than disable it.
  So, enabling the HT or not seems to depend on the software settings, while the TBT no doubt can help shorten the total time cost.
%

\subsection{Possibility of Missing New Pulsars}

  For PMPS data, we only processed 16\% of the data.
  Assuming that the discovery rate is same, we may find 12 new pulsars in total.
  Comparing with the PMPS data processing in recent years (e.g.,  \citealt{Knispel+etal+2013, Eatough+etal+2013}),
  one key point for our discovery can be that all the candidates were checked by human beings.
  We will describe these discovery and publish the human checking results in Pan et al. (in prep) soon.

  For the FAST data, the discovered new pulsars have low DM values (less than 100), since the data is from the 290-340 MHz band.
  No doubt that we may miss high DM pulsars.
  The data is only for 52 seconds observation and thus we indeed missed long period pulsars (approximate period longer than 2 seconds).
  Those pulsars were detected by a single pulse search pipeline (Zhu et al. in prep).

  We have carefully tuned the search parameters before we start searching new pulsars in PMPS data or FAST data,
  yet some new pulsars will still be missed.

\subsection{Using other routines}

  The PRESTO routine {\bf prepdata} is simply used for dedispersing data with one DM value.
  We use it in the pipeline.
  Another routine, {\bf prepsubband} can be used for dedispersing data with a set of DM values.
  We also tested it and compared the time cost.
  When we use it for dedispersing a data file with relatively small size (e.g., less than 10 GB) and then run pulsar search in parallel,
  it is faster or even much faster than using {\bf prepdata} in parallel.
  This is due to saving time from reading data files.
  The situation is similar when using the MPI routine of {\bf prepsubband}, {\bf mpiprepsubband}.

\section{Conclusions}
\label{sect:Conclusions}
  In order to search for pulsars from FAST drift scan data (many small files, only CPU available), we presented the RPPPS package.
  Our conclusions are as follows:


  1) with RPPPS package, we successfully processed FAST ultra-wideband pulsar search data in realtime with 20 computing nodes (480 cores in total).
      The RPPPS package is suitable for processing small files when only the CPUs are available for computing, and can achieve a high efficiency.

  2) till now, two new pulsars from PMPS data and tens of new pulsars from FAST drift scan data have already been discovered with RPPPS.
     The two new pulsars from PMPS data are very faint,
     indicating that the RPPPS can be suitable for finding faint pulsars, including faint new pulsars missed by previous PMPS data processing.
     the new pulsars from FAST data have highly varied parameters (or properties).


\section*{Acknowledgements}

We thank Scott Ransom and anonymous reviewer for useful comments and suggestions. This work is supported by National Key R\&D Program of China No. 2017YFA0402600, State Key Development Program for Basic Research (2015CB857100),
National Basic Research Program of China (973 program) No.2015CB857101,
National Natural Science Foundation of China under Grand No. 11703047, 11773041, U1631132 and U1831131, ZP is supported by the CAS ¡°Light of West China¡± Program, 
Supported by CAS Key Laboratory of FAST, National Astronomical Observatories, Chinese Academy of Sciences, Beijing 100101, China.
This work made use of data from the Five-hundred-meter Aperture Spherical radio Telescope (FAST),
FAST is a Chinese national mega-science facility, built and operated by the National Astronomical Observatories, Chinese Academy of Sciences.












\label{lastpage}

\end{document}